 \documentstyle[prl,epsfig,aps]{revtex}

\begin{document}
\draft

%
  \wideabs{
%

\title{First-principles calculations of the self-trapped exciton in 
crystalline NaCl}

\author{Vasili Perebeinos and Philip B. Allen}
\address{Department of Physics and Astronomy, State University of New York,
Stony Brook, NY 11794-3800}
\author{M. Weinert}
\address{Department of Physics, Brookhaven National Laboratory, Upton, 
NY 11973-5000}
\date{\today}
\maketitle

\begin{abstract}
The atomic and electronic structure of the lowest triplet state of the
off-center (C$_{\text{2v}}$ symmetry) self-trapped exciton (STE) in
crystalline NaCl is calculated using the local-spin-density (LSDA)
approximation. In addition, the Franck-Condon broadening of the
luminescence peak and the a$_{1g}\rightarrow\text{b}_{3u}$ absorption
peak are calculated and compared to experiment.  LSDA accurately
predicts transition energies if the initial and final states are both
localized or delocalized, but 1 eV discrepancies with experiment occur
if one state is localized and the other is delocalized.
\end{abstract}

\pacs{71.35.Aa, 61.82.Ms, 78.20.Bh, 71.20.Ps}

%
  }   
%


Unlike a molecule, an extended system such as a solid can support both 
spatially localized and delocalized single particle states and excitations.
Physical properties such as luminescence can differ dramatically depending 
on which type of state occurs. Deciding theoretically whether a localized 
or delocalized solution exists is a challenging problem 
\cite{Mauri}. Here we examine NaCl, a classic example \cite{Williams1} 
where electronic excitations self-localize by coupling to the lattice,
creating local lattice distortions. Because the degree of localization
will affect the Coulomb energies,
approaches
that incompletely cancel the 
self-interaction contribution to the exchange energy
(e.g., the local density approximation)
sometimes fail to predict the actual localized solution.

The ground state of alkali halides with one electron removed is the
V$_K$ center \cite{Williams1,Castner}: the resulting hole does not 
delocalize at the top of the valence band, but rather (symmetrically) 
attracts two
Cl$^-$ ions into a tightly bound molecule \cite{Kabler,Slichter},
effectively becoming a Cl$_2^-$ molecular ion. The local symmetry of
this atomic configuration is D$_{\text{2h}}$.  An excess electron in
bulk NaCl forms a large mobile (Fr\"{o}hlich) polaron \cite{Devreese}, 
but in the
presence of a self-trapped hole forms a self-trapped exciton (STE). It
has been suggested \cite{Williams2} that the self-trapped exciton state
breaks symmetry and sits off-center with C$_{\text{2v}}$ symmetry.

Although there have been many theoretical studies
\cite{Williams2,Jette,Shluger,Puchin2,Derenzo} of the self-trapped
exciton and V$_K$ center problems, no density-functional calculations
have been reported.  In this letter we report local-spin-density
approximation (LSDA) calculations of the STE and V$_K$ center. The
advantage of LSDA calculations is that they provide one of the simplest
tools capable of providing a realistic model of this competition.  In
both cases, our LSDA calculations give undistorted, delocalized
solutions with lower energy than the self-trapped solution, contrary to
experiment.  For the V$_K$ center, no metastable local minimum trapped
solution was found; however, for the (neutral) STE we find locally {\em
metastable\/} solutions, with the on-center STE 0.14 eV higher in
energy than the off-center STE solution, which in turn is higher by
0.20 eV than a free electron-hole pair. Even with this discrepancy in
the total energy,
the atomic positions for
the STE solution (see Table \ref{tab1}) are reasonable and agree well
with Hartree-Fock second-order M{\o}ller-Plesset (MP2) perturbation
theory \cite{Puchin2}. We focus on the properties of the local
minimum solution for the off-center STE, which provides a test of the
ability of density functional methods to treat the strong coupling of
electronic and lattice degrees of freedom, and calculate the spectral
properties of excited states.

\begin{figure}
\centerline{
\psfig{figure=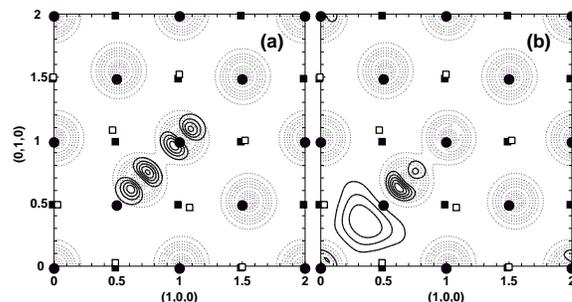,height=1.55in,width=2.95in,angle=0}}
\caption{The dotted contours show the total valence charge density  
in the STE. The solid contours show $|\Psi|^2$ for the trapped 
(a) hole and (b) electron.
The open (filled) squares represent displaced (ideal) positions 
of the Na atoms; filled circles represent undisplaced Cl atoms.
The lowest contours are 0.04 for the 
total charge, 0.01 (0.002) for the hole (electron) state,
with increments of 0.05, 0.01 and 0.002 
e (a.u.)$^{-3}$, respectively.}
\label{rho}
\end{figure}

To solve the
LSDA equations we use a plane wave pseudopotential method
\cite{Chetty,Calcs1,Tests1}
with a spin-dependent exchange-correlation potential
\cite{Ceperley}, and full structural relaxation in a supercell approach.
For most calculations, we used a 32 atom supercell with translation
vectors (2,0,0), (0,2,0) and (1,1,1), giving a nearest neighbor
distance between STEs of 9.4 {\AA}, and used four special
{\bf k}-points in the irreducible wedge for the
Brillouin zone integrations. Tests varying the number of {\bf k}-points
and supercell size suggest that these parameters are adequate
\cite{Tests2}.
One of the in $x-y$ plane nearest Cl and two Na atoms are most displaced while 
the rest of the atoms move by a much 
smaller amount. Some calculated structural parameters for
the STE are given in Table \ref{tab1}.

The calculated atomic displacements from the ideal NaCl structure for
the off-center STE are shown in Fig.\ \ref{rho}.  The  b$_{3u}$ hole
state in the triplet STE is localized on the Cl$_2^-$ molecule
(Fig.\ \ref{rho}a), with nearly equal weight on the
two Cl ions.  The last spin up electron (a$_{1g}$) is mostly
localized on the (1/2,1/2,0) vacant halogen site (Fig.\ \ref{rho}b), as
in the case of the F-center.  The formation of the
Cl$^-_2$ ``molecule'' in the STE is mainly due to the shift of a single
Cl. This asymmetric shift can be rationalized by noting that the
Madelung energy (with canonical charges of $\pm$1 for Na and Cl) of
the D$_{\text{2h}}$ configuration is about 0.17 eV higher than the
C$_{\text{2v}}$ one, i.e., the ionic Madelung terms that favor
the rock salt structure in the first place favor keeping one of the Cl
ions on a lattice site; in addition, the extra electron in the STE
(compared to the $V_K$ center) can lower its energy by this distortion.

\begin{figure}
\psfig{figure=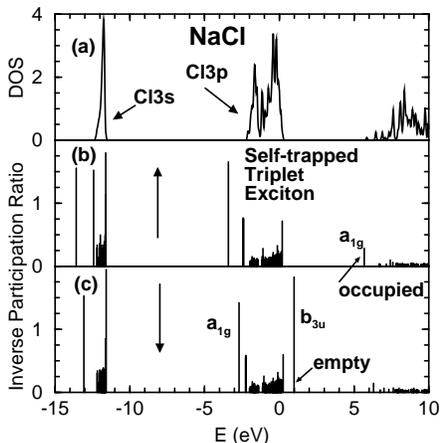,height=2.3in,angle=0}
\caption{(a) density of states (states/eV spin formula unit) of pure NaCl 
crystal, where the two spin orientations are equivalent. IPR of triplet 
STE for (b) spin up and (c) spin down electrons.
The state labeled b$_{3u}$ is empty and the state a$_{1g}$ is occupied in 
the STE. The D$_{\text{2h}}$ representations b$_{3u}$ and a$_{1g}$ became 
indistinguishable (both A$_1$) in the true C$_{\text{2v}}$ symmetry of the 
off-center state.}
\label{dos}
\end{figure}

To obtain vibrational properties of the off-center STE, we made finite
displacements from the equilibrium geometry. The Cl$_2^-$ stretching
mode $\omega_{\text{str}}=242\,\text{cm}^{-1}$ is smaller than the
experimental Raman frequency 361 cm$^{-1}$ \cite{Tanimura1}.
>From the force matrix associated with the Cl$_2^-$ stretching mode,  we
found that only the two neighboring Na ions at ($\frac{1}{2}$,1,0) and
(1,$\frac{1}{2}$,0) couple significantly.  Unlike
H center calculations \cite{Svane1}, coupling to the Na atoms yields
only a small shift in the frequency of the  Cl$_2^-$ mode to
$\omega_{\text{str}}=234\,\text{cm}^{-1}$.  Since the LSDA places the
STE too high in energy, it
is not too surprising that the curvature of the STE
local minimum is underestimated,

Figure \ref{dos} shows the computed density of states (DOS) of the
perfect NaCl and the inverse participation
ratio (IPR=$\text{a}_0^3/8\int\ |\Psi_{i}(\vec{r})|^4d^3r$) for states
of majority and minority spin for the off-center STE. (The IPR is a
measure of the localization of a state.) The localized hole b$_{3u}$
and electron a$_{1g}$ states lie in a gap of about 6 eV between the
conduction and valence bands of the perfect crystal, where as usual the
LSDA underestimates the gap.  Rather than using the single-particle
eigenvalues, we obtain estimates of the excitation energies as the
difference between total energies of different electronic
configurations.  To calculate the energy to create a free electron-hole
pair (the gap energy), we occupy spin up states with one extra
electron, while spin down states have one empty state.  The energy
difference between the two solutions (6.44 eV) for the same  atomic
positions should correspond to the free electron-hole pair
(experimental value 7.96 eV \cite{Knox}).

\begin{figure}
\psfig{figure=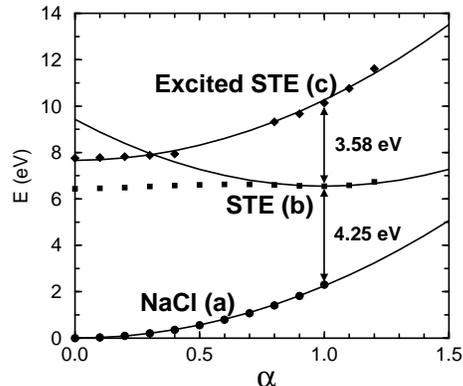,height=2.0in,width=2.35in,angle=0}
\caption{Potential curves for pure NaCl (a), 
self-trapped exciton in the 
ground state (b) and electronically excited state 
(a$_{1g}\rightarrow$ b$_{3u}$) (c).
The coordinate $\alpha$ measures the magnitude 
of the displacement, such that $\alpha=0$ and $\alpha=1$ correspond to the 
undisplaced NaCl crystal and the relaxed STE atomic configurations,
respectively.}
\label{pot}
\end{figure}

In the distorted STE solution, the energy of electron-hole pair
recombination was found by comparing energies of the STE solution with
the energy of NaCl having the same atomic displacements as that of the
STE. This energy of 4.25 eV is roughly the same as the energy difference 
between a$_{1g}$ and b$_{3u}$ states given by the DOS, and compares to a 
value of 3.35 eV  obtained from a luminescence experiment \cite{Ikezawa} 
at 11 K. The distorted NaCl, with a lattice distortion energy
of 2.5 eV relative to the ideal NaCl positions, is in a highly
excited vibrational state. Thus, significant
Franck-Condon effects in the spectral properties of the STE are expected.  To
qualitatively describe the luminescence, a ground state potential curve was
calculated for the configuration coordinate $\alpha$.  It was assumed
that all atoms move back to the perfect crystal positions
proportionally to their distortions in the STE solution. The result
is shown in Fig.\ \ref{pot}, along with a
quadratic fit $k_a\alpha^2/2$ with one adjustable parameter $k_a$. The
effective one-dimensional Schr\"odinger equation for $\Psi(\alpha)$
describes a harmonic oscillator with frequency
$\omega_a=\sqrt{k_a/I}=109 \,\text{cm}^{-1}$.  The moment of inertia
$I=\sum{M_n\delta\vec{R}_n^2}$ was chosen so that $I\dot{\alpha}^2/2$
equals the kinetic energy of the atoms with mass $M_n$, when they move
from the initial displacements $\delta\vec{R_n}$.

The same type of the potential energy curve was calculated for the STE.
The quadratic fit works only in the close vicinity of the exciton
{\em metastable\/} minimum.  The resulting frequency is 
$\omega_b=123\,\text{cm}^{-1}$.
The experimental \cite{Ikezawa} temperature T=11 K justifies a zero
temperature approximation (the STE initial state is the vibrational
ground state).
Since the luminescence peak position corresponds to the vibrational
level $n$$\approx$170 of the electronic ground state, quasiclassical
wavefunctions were used in the numerical integral evaluation.  The
sequence of vibrational sidebands should be replaced by a sequence of
convolved densities of phonon states $D(\omega)$.  We approximate this
by a Gaussian,
$D(\omega)\rightarrow\exp(-\omega^2/2\gamma^2)/\sqrt{2\pi}\gamma$ with
the width $\gamma$=43 cm$^{-1}$ chosen such that the first three
moments coincide with the experimental \cite{Raunio} phonon DOS. This
gives a luminescence width of 0.43 eV, while the experimental width is
0.63 eV.

\begin{figure}
\psfig{figure=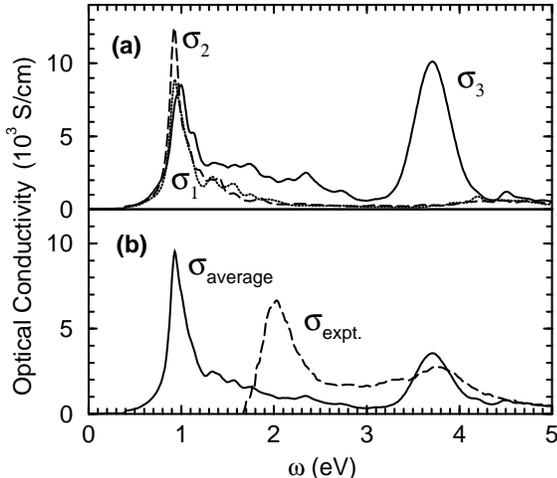,height=2.5in,width=2.9in,angle=0}
\caption{Optical conductivity of self-trapped exciton in NaCl normalized per 
volume of a single exciton $a_0^3/2$. 
(a) $\sigma_1(\omega)$: $\vec{E}\bot$ to the molecular
axis in the $x-y$ plane; $\sigma_2(\omega)$: $\vec{E}$ along the $z$ 
direction; and $\sigma_3(\omega)$: $\vec{E}$ along the Cl$_2^-$ molecule.
(b): average conductivity compared with experiment 
\protect\cite{Williams3}.}
\label{opt}
\end{figure}

The optical response $\sigma(\omega)$ of the long-lived triplet STE 
also has been measured \cite{Williams3,Tanimura2}.
The diagonal part of the optical conductivity tensor is:
\begin{eqnarray}
\sigma_{\alpha}(\omega)=\frac{\pi e^2 N}{m^2\omega\Omega}\sum_{\ell \ne \ell'}
\sum_{\vec{k}} (f_{\ell \vec{k}}-f_{\ell'\vec{k}})
\nonumber \\
\mid \langle \ell\vec{k}
|p_{\alpha}|\ell'\vec{k}\rangle\mid^2
\delta(\hbar\omega-E_{\ell'\vec{k}}+
E_{\ell\vec{k}})
\label{sigma}
\end{eqnarray}
where $f_{\ell \vec{k}}$ is the occupancy of the state $|\ell\vec{k}\rangle$.
The spin state index is included in the band index $\ell$.  Integration
over the zone has been performed using 26 {\bf k}-points in the
irreducible zone.
In Fig.\ \ref{opt}a, absolute optical conductivity 
curves are shown for the three 
polarizations, $\vec{E}\parallel$(1,-1,0), (0,0,1), and 
(1,1,0) (parallel to the the Cl$_2^-$ molecular axis).
Fig.\ \ref{opt}b shows the average conductivity 
$\sigma(\omega)=\sum_i\sigma_i(\omega)/3$, which
is compared with experiment \cite{Williams3,Tanimura2}, 
rescaled so that the total weights under the both curves are the same.

The first peak in Fig. \ref{opt}b is centered at 0.95 eV; the
splitting between the peaks for the three different polarizations is
not resolved in the calculations.  Most of the weight in these
peaks comes from transitions of the last localized spin-up a$_{1g}$
electron into empty conduction band delocalized states (see Fig.\
\ref{dos}b).

The second peak centered at 3.58 eV is the
a$_{1g}\rightarrow\text{b}_{3u}$ transition for the spin down electron
(see Fig. \ref{dos}c).  The energy difference between the ground
state and electronically excited exciton state with the same atomic
configuration turns out to be the same as the eigenvalue difference of
b$_{3u}$ and a$_{1g}$ states of the ground STE.  The excited exciton
will lower its energy by moving atoms back to the undistorted positions
of perfect NaCl. To apply the Franck-Condon principle, we repeat the
same type of calculations for the excited exciton as we did for NaCl
and the ground state STE (Fig. \ref{pot}). When two Cl atoms move away
from each other, the a$_{1g}$ empty state merges with the valence Cl 3$p$
band, which makes it very difficult to choose which
state to depopulate during the iterations.  Instead we used the results for the ground state
STE to obtain the energy of the electronically excited state by adding
eigenvalue difference $\lambda_{\ell 1}-\lambda_{\ell 2}$ between two
states for which dipole matrix element $\mid\langle\ell1|p_3|\ell
2\rangle\mid^2$ is the largest.  Results are shown on Fig. \ref{pot}
along with a quadratic fit ($\omega_c$=116 cm$^{-1}$) which works
for the entire range of the parameter $\alpha$.  The delta-function of
Eq.\ (\ref{sigma}) corresponding to the a$_{1g}$$\rightarrow$b$_{3u}$
transition was replaced by a sequence of convolved Gaussian peaks.

The transition energy (3.58 eV) between the two localized states
a$_{1g}$$\rightarrow$b$_{3u}$ (see Fig.\ \ref{dos}c) agrees well
with the experimental peak at 3.8 eV.  But when initial and final
states have different degrees of localization, errors of an eV in the
luminescence (LSDA: 4.25 eV, expt. \cite{Ikezawa}:  3.35 eV) and in the
optical excitation of the bound electron into the conduction band
(LSDA: 0.95 eV, expt. \cite{Tanimura2}: 1.95, 2.13 and 2.00 eV for
three different field polarizations) occur. 
A possible explanation for this error is that in LSDA
the incomplete
cancellation of the large repulsive
self-interaction energy 
$\int d\vec{r}d\vec{r'} n_i(\vec{r})n_i(\vec{r'})/|\vec{r}-\vec{r'}|$
of a localized state $n_i=|\Psi_i|^2$ and
the corresponding exchange term may
destabilize a localized solution in favor of a delocalized 
one. Corrections such as
self-interaction corrections (SIC) \cite{Perdew} or LDA+U 
\cite{Liechtenstein} may reduce the error; in SIC,
shifts in the energy of a state $i$ on the order of $\int
d^3r n_{i}^{4/3}(\vec{r})$ \cite{Perdew} are expected. If the degrees of
localization of the two states are similar, however, then the LSDA
transition energies are reasonable.

In summary, we have presented LSDA calculations for the STE in NaCl,
including the coupling between the lattice and electronic states. The
off-center STE is found to be more stable than the on-center STE, but
both are metastable compared to free electron-hole pairs.  Both
luminescence and optical conductivity, including vibrational
Franck-Condon effects, were also calculated. The hole state in the
off-center STE is found to be rather evenly split between the two Cl
atoms, but the electron state is localized to the vacant site left by the 
shifted Cl ion.  The
density functional description of electronic transitions between
localized states, such as the a$_{1g}\rightarrow\text{b}_{3u}$
absorption peak, agree well with experiment.  For transitions between
the localized and delocalized states, discrepancies of order 1 eV with
experiment arise.  Although the LSDA can capture many features of the
STE states, when a localized solution competes with a delocalized
solution, the incomplete cancellation of the self-interaction may
destabilize the localized solution. Given the usefulness of the LSDA 
method for unraveling complex materials, it is important to test and 
develop approaches that can treat localized and delocalized states on 
the same footing.

\acknowledgements
We thank M. L. Cohen, G. W. Fernando,  P. M. Johnson, S.G. Louie, 
W. E. Pickett for discussions.  
This work was supported in part by NSF Grant No.\ DMR-9725037 and by DOE
Grant No.\ DE-AC-02-98CH10886.

\begin{table}
\caption{The equilibrium Cl-Cl distance $r_{Cl-Cl}$, 
displacements of the nearest Na ions $\Delta_{Na}$ in ({\AA}), 
for the STE. Present results are compared to 
the Hartree-Fock (MP2) theory \protect\cite{Puchin2}.} 
\label{tab1}
\begin{tabular}{ccc}
Configuration & $r_{Cl-Cl}$(MP2) & $\Delta_{Na}$(MP2) \\
\hline
STE D$_{\text{2h}}$ & 2.73 (2.654) & 0.35 (0.383) \\
STE C$_{\text{2v}}$ & 2.59 (2.525)  & 0.51 (0.488)  \\ 
\end{tabular}
\end{table}

\end{document}